# Tuning InP self-assembled quantum structures to telecom wavelength: a versatile original InP(As) nanostructure "workshop".

E. E. Mura,[1,a] A. Gocalinska,[1] G. Juska,[1] S. T. Moroni,[1] A. Pescaglini,[1] and E. Pelucchi,[1]

[1] *Epitaxy and Physics of Nanostructures*, Tyndall National Institute, University College Cork, "Lee Maltings", Dyke Parade, Cork, Ireland

*Abstract*

The influence of hydride exposure on previously unreported self-assembled InP(As) nanostructures is investigated, showing an unexpected morphological variability with growth parameters, and producing a large family of InP(As) nanostructures by metalorganic vapour phase epitaxy, from dome and ring-like structures to double dot in a ring ensembles. Moreover, preliminary microphotoluminescence data are indicating the capped rings system as an interesting candidate for single quantum emitters at telecom wavelengths, potentially becoming a possible alternative to InAs QDs for quantum technology and telecom applications.

In the past few decades low-dimensional self-assembled quantum dots (QDs), as well as other zero/one dimensional nanostructures, have played an important role in studies of fundamental and applied physics[1], being attractive, for example, for optoelectronic applications[2,3], as well for quantum information and computing[4,5,6]. The possibility to transition/transform from one nanostructure into another is a topic of particular interest, e.g. a dotlike shape can be made into a ring-like one, i.e. into quantum rings (QRs).[7,8,9,10,11,12,13,14,15,16]

Considering AlInAs–InP structures, conventionally exploited in telecom applications, they show a peculiar type II band alignment, making them suitable for conventional opto-microelectronics[17], while can also be an attractive option for "exotic" innovative applications with engineered band alignment. Some of us have shown unforeseen evidence[18] that the presence of $Al_{0.48}In_{0.52}As$, together with specific surface organization (and possible phase separation), has profound effects on the nucleation of InP (mono)layers. Indeed, InP deposited directly on lattice-matched AlInAs forms a variety of nanostructures, despite the nominally strain free environment. These structures, if capped, would generally result in a relatively broad type II emission around one micron, while preliminary transmission microscopy images show that the nanostructures evolve during capping, and are partially preserved after overgrowth.[19]

In this Letter, we report on the possibility of transforming these InP nanostructures grown by metalorganic vapour phase epitaxy (MOVPE) on lattice-matched InP substrates into variously shaped InP(As) emitters in the telecom windows. For example, InP(As) dots-to-rings transformation is carried out by intentional exposition to selected hydrides. Moreover, we observe a large phenomenology of morphologies induced by growth conditions choice and post-growth layer exposure to hydrides that we critically discuss. Indeed the combination of arsenisation and cooldown protocols affects the final nanostructures' shape, changing the original dots into rings or domes (and others), conveying a rich variety of nanostructures in a controlled manner, and importantly engineering the photoluminescence characteristics, with specific configurations delivering clear signatures of single dot emission at both 1.3 and 1.55 microns.

---

[a] Author to whom correspondence should be addressed. Electronic mail: enrica.mura@tyndall.ie

Sample growths were carried out in a commercial horizontal MOVPE reactor at low pressure (80 millibars) with purified $N_2$ as the carrier gas[20]. The precursors were trimethylindium (TMIn), trimethylaluminum/gallium (TMAl/ TMGa), arsine ($AsH_3$), and phosphine ($PH_3$).

Concerning morphological studies different sample designs were grown. Due to former laboratory practice , and to keep consistency with previous work, a first family of samples (which we will refer as "combined seed heterostructure") comprised of a 100 nm $Al_{0.48}In_{0.52}As$ layer (lattice matched to InP) grown on a 100 nm homoepitaxial InP buffer on an <100> InP semi-insulating substrates, nominally perfectly oriented. Then 20 nm of $In_{0.10}Ga_{0.90}P$ were deposited followed by 0.5 nm of $Al_{0.48}In_{0.52}As$. InP based nanostructures were then grown depositing a thin InP film (1 nm) on the previous 0.5 nm AlInAs layer. Following the InP film deposition, the nanostructures were then transformed by exposure to $AsH_3$ and $PH_3$ as required, then the precursor was switched to $PH_3$ or $AsH_3$ for the cooldown protocol[18] (structure sketch and process diagrams included in Figure 1). In the second set the structure was intentionally simplified removing the 20 nm $In_{0.10}Ga_{0.90}P$ and 0.5 nm of $Al_{0.48}In_{0.52}As$ layers. In this "simple seed-structure" the InP based nanostructures were grown directly on the 100 nm of $Al_{0.48}In_{0.52}As$ layer[18] with the same protocol as in the "combined seed heterostructure", including exposure and cooldown. Indeed we reported results on both for scientific thoroughness (the InGaP layer was historically broadly used in our InP nanostructure work to test the effects of thin AlInAs layers and possible group III adatom exchange. This work started on well characterized structures including the InGaP layers, and only subsequently evolved to a simplified seed structure on which we obtained our best optical results), and on each sample we specified the seed structure of choice. At this stage, in general we did not notice significant differences with seed choice within statistical sample to sample variations, if not when we highlighted them in our text. Atomic force microscopy (AFM) was then used to probe the surface morphology, in tapping/non contact mode at room temperature and in air.

Most samples for photoluminescence (PL) measurements shared one of the structures discussed before and were than capped with additional 20 nm $In_{0.10}Ga_{0.90}P$ and/or 100 nm $Al_{0.48}In_{0.52}As$ grown on top of the "arsenised" nanostructures, as discussed in detail in the text.

For all samples the InP buffer growth conditions were as in Ref. 21 for consistency. Growth conditions for the AlInAs layer were fixed for all the samples: V/III ratio of 120, growth rate Gr = 1 µm/h, real estimated, by emissivity corrected pyrometry, growth temperature $T_g \approx 600$ °C. The growth conditions used here for the growth of 1 nm InP layers were: V/III ratio of 180, Gr = 0.7 µm/hr, $T_g \approx 630$ °C. The self-organized InP nanostructures growth by MOVPE are described in more detail elsewhere.[18]

X-ray diffraction was used when necessary to confirm macroscopic lattice matching of the $Al_xIn_{1-x}As$ layers. The optical characterization of capped samples was carried out in our micro-photoluminescence spectroscopy setup at cryogenic temperatures, using a laser diode emitting at 633 nm.[22]

In Fig. 1 10x10 µm$^2$ AFM images of representative InP nanostructures are shown (all obtained on "combined seed heterostructure" like designs). To give a "baseline", Fig. 1(a) shows InP QD nanostructures obtained as in Ref. 18 (i.e. by depositing InP directly on lattice matched AlInAs, and critically cooling them down under $PH_3$). Here, a mixture of islands and small rings emerges. The height of the QDs doesn't exceeded 8 nm with a base diameter with a maximum of ~300 nm and areal density of ~1.9x10$^8$ cm$^{-2}$. Nanorings exhibit a base diameter of ~300-200 nm outer and ~200-100 nm inner. The overall height is comparable with the QDs whereas the areal density is slightly lower, about ~1.2x10$^8$ cm$^{-2}$.

In Fig. 1(b) we show the morphology of one sample in which a previously deposited InP layer is subsequently exposed to an arsine flow ($AsH_3$ provided post-growth and during cooldown time, as described in the figure caption). Nanorings disappeared, giving way to large dome-like structures[23]. The areal density, ~2.7x10$^7$cm$^{-2}$, is notably one order of magnitude lower when compared to the previous structure without the arsenisation process. The features are slightly elongated in the [011] direction with lateral dimension of ~430 nm and ~320 nm. The domes are unexpectedly tall, with height in excess of 150 nm. It should also be said that in other samples where the constant growth T arsenisation was kept for longer (5 minutes instead of one), no significant morphological differences were detected by AFM (*not shown*). Altogether, the arsenisation period seems to not only substitute phosphorous atoms in the original structures, but also rearrange adatom distribution, enhancing their attachment to selected seeds, and presumably promote Ostwald ripening-like processes, hence the reduced density and larger overall dimensions.[24, 25]

For the sample shown in Figure 1(c) the final layer was exposed to the same arsine flow (one minute $AsH_3$) as in Fig. 1(b), but it was flushed with $PH_3$ during cooldown. It is evident that the sequence of hydride exposure transmuted the final nanostructures morphology, transforming the original domes into large rings. The rings exhibit inner and outer diameter of ~240 nm and ~640 nm respectively, with areal density similar to the previous sample discussed, ~4.0x10$^7$ cm$^{-2}$ in this image. Some rings acquire irregular shape, appearing elongated in the [011] direction. Similar trends are observed in other reported systems where the anisotropy in the redistribution of the material[26] is ascribed to different diffusion rates of indium atoms along different crystallographic directions[27].

Differently to dome growth (Fig. 1(b)), which seemed relatively insensitive (morphologically speaking) to the duration of the exposure to arsine, the rings as obtained in Fig. 1(c) actually evolve and change shape with arsine interaction time, i.e. the arsine (pre)exposure time is a relevant variable. After five minutes of arsenisation (instead of only one) the structures finally presented a definite and "ordered" round shape (see Fig. 2(a)). It is worth noting that we didn't observe a significant excavation in the centre of the fully formed rings in the investigated samples, coming much closer to an "ideal" ring shape[28] than the typical craterlike InAs/GaAs[29] reported in the literature.

Nevertheless, the rings' shape changed when we grew them on the "simple seed-structure" while keeping the same hydride exposure time of five minutes (Fig. 2(b)). The elimination of the InGaP layer between the AlInAs layer and the InP QDs, and a thicker AlInAs layer as template (instead of only ~2 monolayers) seems to slow down the transformation process from dots to rings, highlighted by a central strip/band inside the ring, with the combined effect to generate what looks like a dot/wire in a ring structure.

It is worth noting that, the heights (~10 nm) and inner diameters of the rings are comparable with those of the dots (those obtained right after InP deposition with no $AsH_3$ flux). In Ref. 18 selective wet chemical etching was shown to indicate that there was a possible difference in composition in the InP dots and the small rings (Fig. 1(a)), with a suspicion of the presence of a compositional gradient in the islands (i.e., the rings and the outer parts of the dots seemed to be more P-alloy like, while the centers of the islands more As-alloy like, at least by the responsivity to the acids used during the etching experiments). This may suggest a transformation process, with the As rich regions in the domes, the ones originally formed during the InP dot nucleation and those obtained by P-As exchange by the successive arsenisation, redistributing during the last phosphorization stage, with the central atoms migrating to the external boundaries.

While this process happens here in a rather striking manner, this is not qualitatively a completely unexpected observation. For example most studies on QR have been done with the typical approach to cover QDs with complete or partial capping layer and post growth annealing processes.[7, 30] Instead, just few growths are reported without the use of any cap layer, for example by annealing as-grown InAs QDs[14] or by direct deposition as in the case of GaSb/GaAs where sometimes As/Sb soaking time is necessary to observe transition between nanostructures[31], or just by changing the amount of deposited GaSb.[11] Most of these studies have been done by molecular beam epitaxy (MBE) and only a few by MOVPE.[9, 12, 32] Anyway in most of these studies strong material rearrangement is observed.

Two main different models have been proposed to explain the ring formation process. In the case of partial QD capping evolution to rings, for example, a thermodynamic model[27, 33] suggest that the presence of the capping layer creates variation in the balance of free energy and induces a force that pulls the QDs structures radially outward, leading to the QRs formation. On the other hand, the second model is based on kinetic considerations, specifically on the different surface diffusion rate of group III atoms. In this work, where QRs are fabricated uncapped, it is unclear which kind of model or combination of both effects take place and bring to QRs formation. Notably in ref. 11 and ref. 14 where samples are fabricated with an MOVPE system, a similar to ours adatom rearrangement is reported; in one case the role of surface As-Sb exchange reactions was identified as one of the factors contributing to the formation of QRs, and in the other case the mechanism is discussed in terms of As/P exchange with a relevant role of strain in the QD/QR system.

We want also to stress that the morphology of the InP(As) nanostructures grown in our InP/AlInAs system is strongly sensitive to hydride exposure and to simple changes in the process parameters. The zoology of morphologies that we observed is indeed broader than what is here reported. Without digressing too much from the core of our contribution, in Fig. 2(c) we show one example: the AFM image of a "simple seed-structure" exposed to five minutes of arsine with the addition of antimony as surfactant. The Sb addition has a relevant effect: a surface organization transformed appears with one or two dots enclosed into elongated rings. The dots and the rings keep roughly the same overall height (~10 nm) and the same outer diameter, respectively, as observed in previous samples. It is worth observing that a similar kind of complexity of morphologies in the III-V system were observed till now only by droplet epitaxy.[34, 35]

The arsenisation of InP nanostructure (correspondent uncapped structures in Fig. 1(b)) changed not only the morphology but, importantly, also the optical properties of the investigated nanostructures. We observed a typical type II band alignment structure emission peaked around 1 micron (Fig. 3) for the InP/AlInAs system in the structures with "pure" InP/AlInAs nanostructures without the "arsenisation" step (Fig. 1(a)) when simply capped with $Al_{0.48}In_{0.52}As$. As an example, the photoluminescence spectrum relative to a "combined seed heterostructure" exposed to <u>seven</u> minutes of Arsine flow (we chose here a longer arsenisation time than previously discussed so to compare to the "optimized" result in Fig. 4) and then capped with 100 nm of $Al_{0.48}In_{0.52}As$ (inset Fig. 3), shows similar features and additional few spread peaks around (1140-1170) nm, with relatively broad "single nanostructure like" emissions (probably linked to the dome structures Fig. 1(b), or better with their evolution with capping).

The spectral characteristic change completely when the "arsenised" nanostructures, grown here following the "simple seed structure", are capped with 20 nm $In_{0.10}Ga_{0.90}P$ and then 100 nm $Al_{0.48}In_{0.52}As$. In Figure 4 we present the low temperature PL spectrum from these capped nanostructures, showing individual transitions with very narrow linewidths (the best one found was 27 µeV in FWHM), and with power dependencies characteristic typical of single quantum dots.[36] They emit in a very attractive (and extraordinarily broad) spectral region, covering nearly the whole 1.1-1.6 micron range (not all shown in Fig. 4). This is very hard to achieve with traditional SK type dots, grown either on GaAs or InP, where emission is concentrated in relatively narrow bands specific to the growth protocol exploited. Indeed the emission from the dots assembly is spread across over more than 350 nm (240 meV) through the telecom window.

Obviously we can only speculate on the complexity of the processes involved, and not exclude some group III rearrangement too as an extra process, not only regarding these specific nanostructures. Indeed the AFM evidence is only obtained after a full sample cooldown, and the nanostructure formation will be forcibly the result of the full process. When samples are capped, the morphological evolution is probably arrested/modified, and clearly different capping strategies affect this evolution and final results, with a phosphorous containing alloy appearing beneficial for obtaining interesting photoluminescence properties. In this respect we present a preliminary investigation in Fig. 5 of AFM images of sample grown in the "simple seed structures", capped with different cap layer thickness. The rings still visible and well defined with 2 nm of InGaP cap change their appearance to volcano-like structures with just 5 nm of InGaP (*not shown*). The surface appears covered with "bubbles" after the deposition of 20 nm of InGaP, keeping the same bumps morphology when 100 nm of AlInAs are added as final layer. On the other hand the rings are replaced by domes when AlInAs is the only cap layer (*not shown*). So in our preliminary tests, the capping InGaP layer was indeed inserted/intended to allow for keeping the ring-shape (and preventing the transformation to domes) and the AlInAs to guarantee a sufficient quantum confinement.

Indeed our results are in early stages, and a more systematic characterization work (outside the limited scope of this report), including extensive transmission electron microscopy images and a broader set of hydrides exposures and capping protocols will be necessary to elucidate and correlate the extraordinary variety of morphological and optical properties here observed. Moreover, selective etching studies on the nanostructures like the ones shown in Ref. 37 are likely to give important insights in the formation mechanism.

In conclusion, the influence of hydrides on unusual self-assembled InPAs nanostructures was investigated, showing an unexpected morphological variability producing a different family of possibly pseudomorphic quantum structures. Notably, we have demonstrated that InP(As) ring-like structures can be spontaneously formed by MOVPE on lattice matched AlInAs. Ring formation is observed when unstrained InP nanostructures are exposed to $AsH_3/PH_3$. Moreover, preliminary microphotoluminescence data are indicating that the capped rings system is an interesting and promising candidate for single quantum emitters at telecom wavelengths, covering a very wide spectral range and delivering narrow emission lines, potentially becoming a possible alternative to InAs QDs for telecom and quantum technology applications.

**Acknowledgements**

This research was enabled by the Irish Higher Education Authority Program for Research in Third Level Institutions (2007-2011) via the INSPIRE Programme and partly by Science Foundation Ireland under the IPIC award 12/RC/2276 and grant 10/IN.1/I3000 and the Irish Research Council under grant EPSPG/2014/35. We thank Kevin Thomas for the MOVPE support.

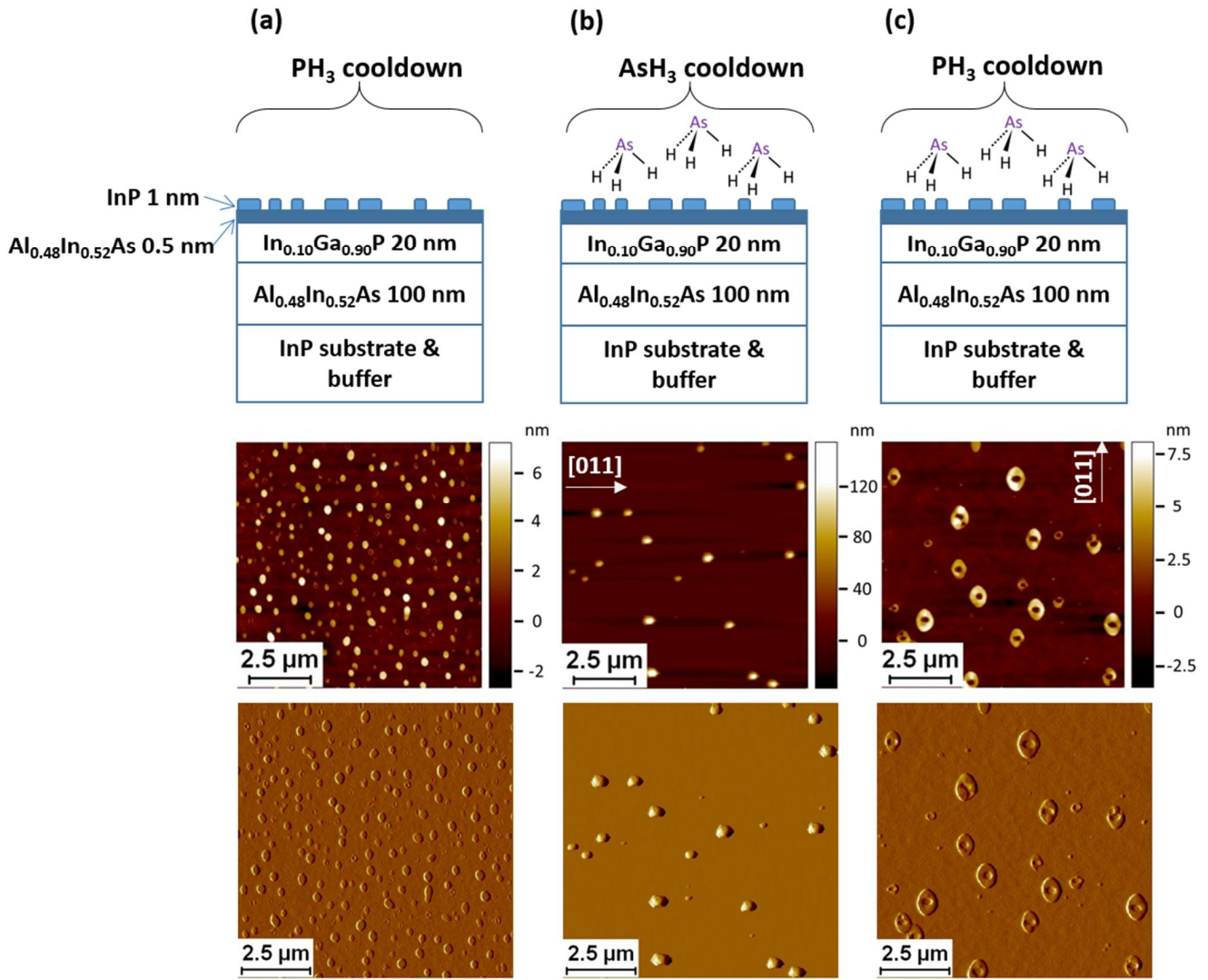

FIG. 1. Structure design and flattened AFM images (height and amplitude signal) of representative InP nanostructures, grown on a "combined seed heterostructure" (a) immediately cooled down under $PH_3$, (b) exposed to arsine at growth temperature for 1 minute at constant T (the same for the InP dot growth) and then flushed with arsine during cooldown, (c) exposed to arsine at growth temperature for 1 minute and then flushed with $PH_3$ during cooldown.

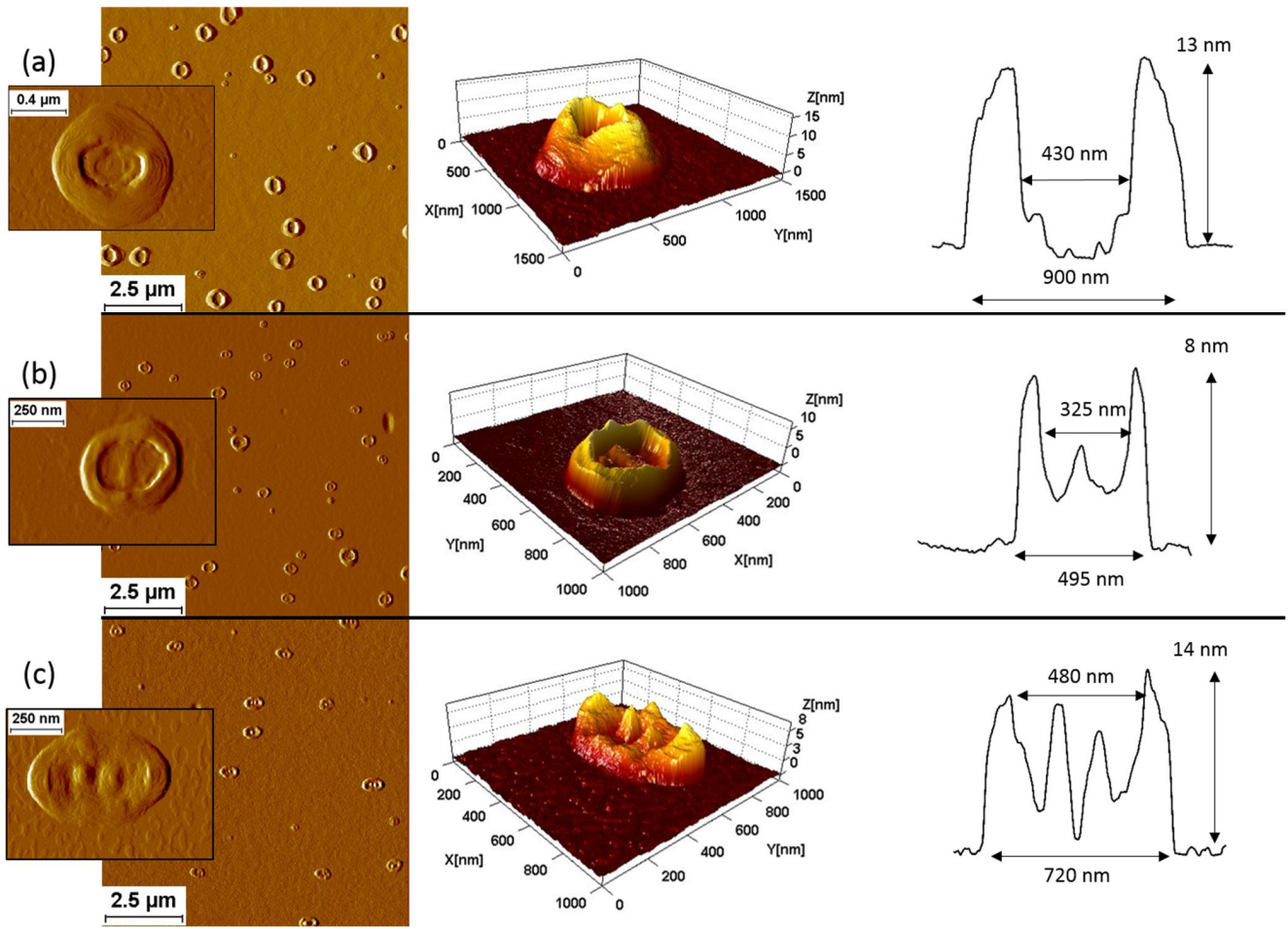

FIG. 2. AFM images (amplitude signal, 3D reconstruction and cross-sectional profile) of InP(As) nanostructures fabricated with same $AsH_3$ exposition time and same cooldown protocol under $PH_3$, but grown on: (a)"combined seed heterostructure" ; (b) "simple seed-structure"; (c) "simple seed structure" exposed to arsine and antimony at growth temperature.

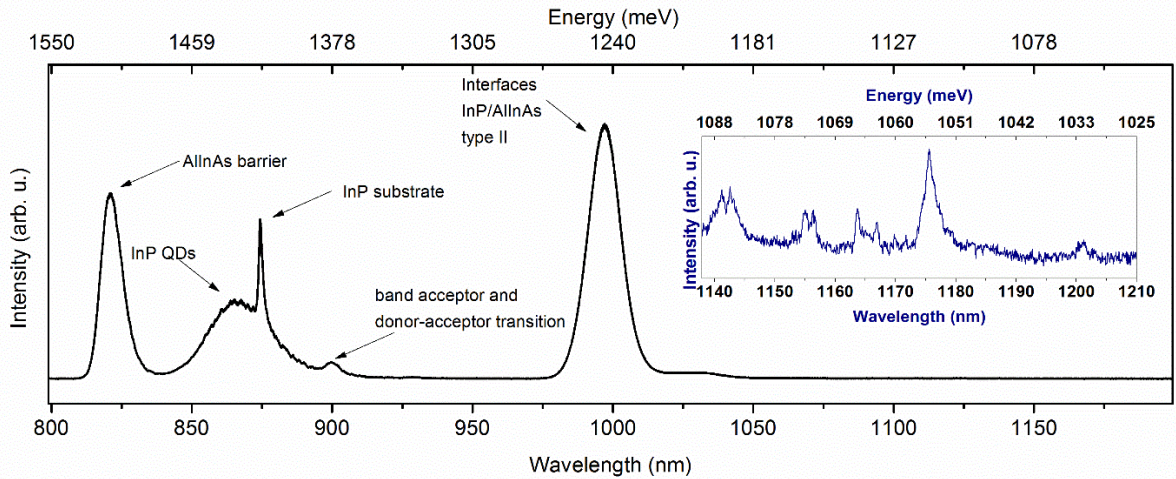

FIG. 3. Low temperature photoluminescence spectrum of a "simple seed structure" directly capped with 300 nm of $Al_{0.48}In_{0.52}As$ and no special hydride treatment. The inset shows a part of photoluminescence spectrum of the "combined seed heterostructure" expose to seven minute of $AsH_3$ and then capped with 100 nm of $Al_{0.48}In_{0.52}As$.

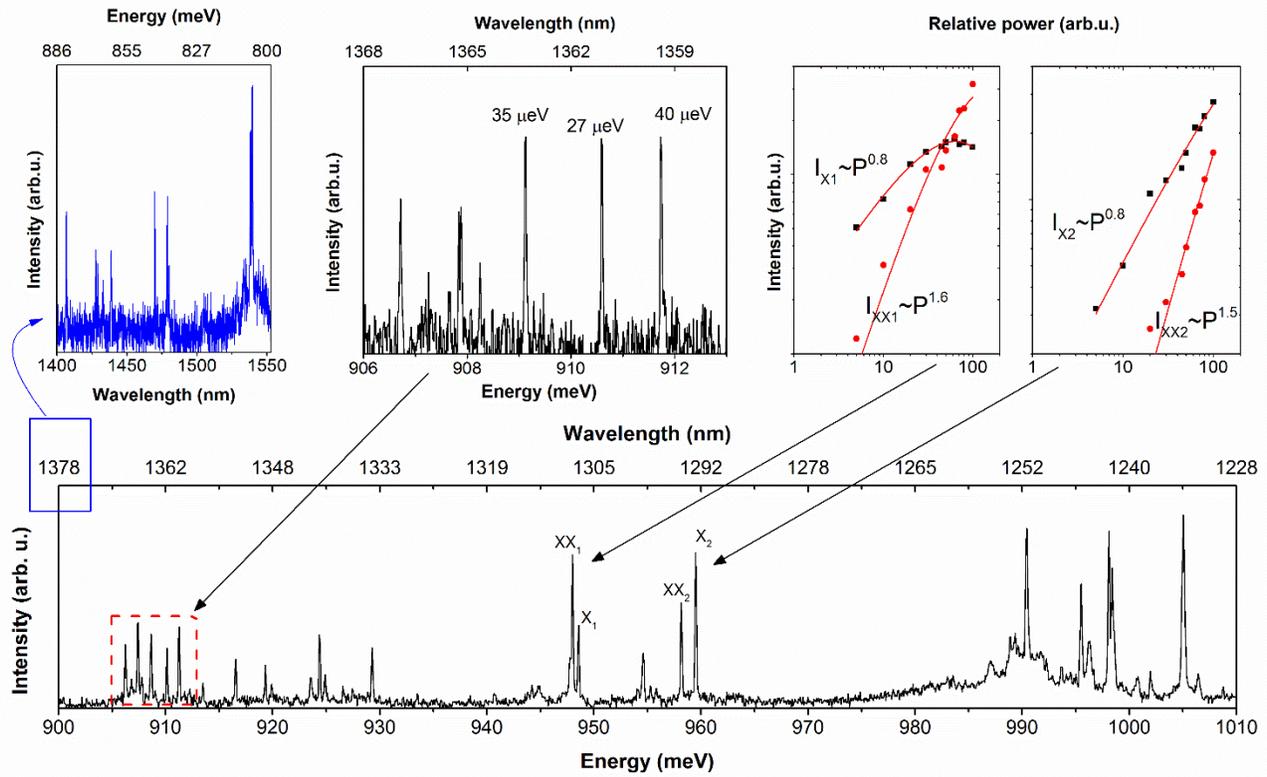

FIG. 4. Part of the low temperature photoluminescence spectrum of "simple seed structure" expose to seven minute of AsH$_3$ and then capped with 20 nm of In$_{0.9}$Ga$_{0.1}$P and 100 nm of Al$_{0.48}$In$_{0.52}$As. Left insert shows detail into a specific spectrum range. Top central insert shows zoom in to the spectrum range with FWHM of transitions stated for each line, top right inserts show power dependence of the peak intensity, allowing for identification of the individual peaks as corresponding to exciton (X) and biexciton (XX) transitions.

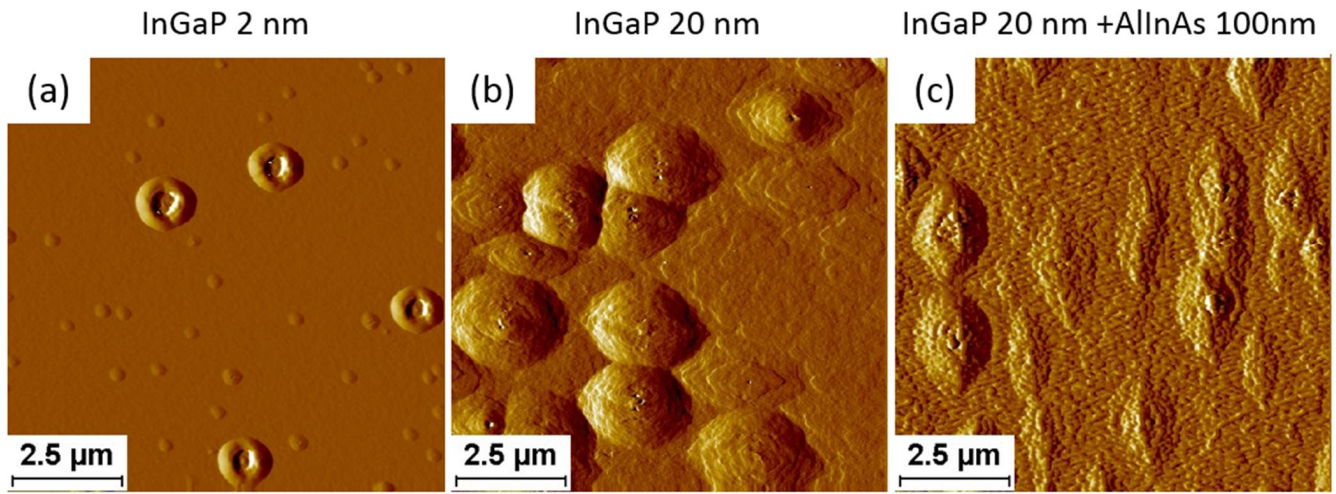

FIG. 5. Flattened AFM amplitude images of samples grown in the "simple seed structure", exposed to arsine flow at growth temperature for 5(a), (b) and 7 minutes (c) and then capped with different cap layer thickness.